\documentclass[aps,prl,twocolumn,superscriptaddres,showpacs]{revtex4}

\usepackage{graphicx}
\usepackage{epsfig}
\usepackage{amsmath, amssymb}

\begin{document}

\title{Dissipationless Information Erasure and Landauer's Principle}
\author{M. Maissam Barkeshli\footnote{Present Address: \it Department of Physics, Massachusetts Institute of Technology, Cambridge, MA 02139}}
\affiliation{Department of Physics, University of California, Berkeley, CA 94720}
\affiliation{Department of Electrical Engineering and Computer Science, 
University of California, Berkeley, CA 94720}

\begin{abstract}

It is widely accepted that information erasure entails heat dissipation. Here we analyze
asymmetric memory states to show that this energy cost can be shuffled around to any step in a 
write-erase cycle and need not accompany the logically irreversible step in a computation. 
We discuss the resulting symmetry between logically irreversible and reversible operations, along with the implications for resolutions of Maxwell's demon paradox.

\end{abstract}


\maketitle

Landauer argued that the act of destroying information during a logical operation
carries with it an unavoidable dissipation of heat \cite{Landauer61}. However, he only explicitly considered bits with
a symmetric physical structure (Fig. 1a) and claimed that the basic physical results hold more generally.
Consequently, it is commonly held that there exists the following
fundamental physical distinction between logically irreversible operations
and logically reversible ones: the former are always accompanied 
by an increase of entropy in the environment while the latter are dissipationless in principle. 

In this letter, we prove that logically irreversible operations -- that is, operations that erase information -- can
indeed be dissipationless in principle. We consider 
asymmetric bits (e.g. Fig. 1b) to show that information erasure exhibits an energy cost only in a complete computational cycle.
All logically irreversible operations can be performed without increasing the entropy of the surrounding environment, provided
that they change the average number of \sc one\rm s and \sc zero\rm s in the bits on which they act. 

The crucial observation is that while information erasure may entail
a decrease in the Shannon entropy \cite{Shannon} of the system representing the information, additional asymmetries can be 
incorporated into the physical structure of the bits to contain this entropy change entirely within
the system itself. These asymmetries will in turn necessitate heat dissipation during a logical
operation that returns the bits involved in the computation to their initial statistics. 

\begin{figure}[h]
\includegraphics[width=3.3in]{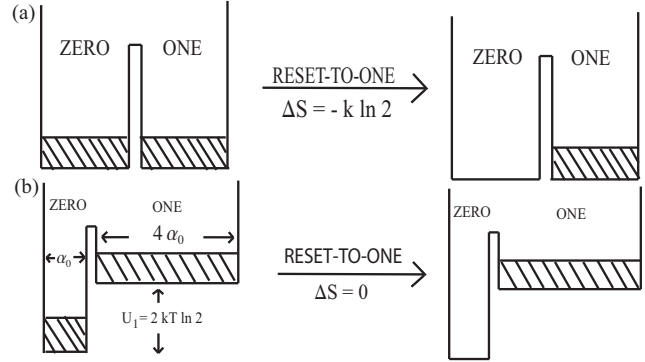}
\caption{Potential energy diagrams for the information-bearing degree of freedom of a bit. Diagonal lines
represent the possibility of finding a randomly chosen bit in various places in the 
bistable potential well. Before the reset operation, \sc zero \rm and \sc one \rm are occupied with equal probability.
}
\end{figure}

The traditional argument for Landauer's ideas is an intuitive one: the erasure of information
takes a bit that could have been in two distinct states, \sc zero \rm and \sc one\rm, and sets it
to a single predefined state regardless of its initial value. This decreases the number of possible
physical states that the bit can occupy by roughly a half, implying a decrease in the 
entropy of the physical system representing the information by $k\ln2$ per bit, where $k$ is Boltzmann's constant.

To be precise, let us represent a bit of information as the 
position of a particle in a one-dimensional bistable potential that is permanently coupled to a
thermal bath at temperature $T$. If the particle is in the region $x < 0$, we will say that the bit is in
the \sc zero \rm state, otherwise it is in the \sc one \rm
state. A practical bit must be static, thus the potential barrier between the two minima must be much larger than $kT$ and so we
will assume that each bit is locally in thermal equilibrium but cannot
cross the barrier on time-scales that are of interest to us. If we
consider a set of bits where each bit is \sc zero \rm
with probability $p_0^s$ and \sc one \rm with probability $p_1^s$,
then the probability that a particle in a randomly chosen bit is at position $x$ is given by

\begin{equation}
\rho(x) = \left\{ \begin{array}{cc}
                   p_0^s \rho_0(x) & \mbox{for $x \leq 0$} \\
                   p_1^s \rho_1(x) & \mbox{for $x \geq 0$}
                   \end{array}
          \right. .
\end{equation}
$\rho_0(x) \simeq \frac{e^{-\beta U(x)}}{\int_{-\infty}^0 e^{-\beta U(x)}
dx}$ and $\rho_1(x) \simeq \frac{e^{-\beta U(x)}}{\int_{0}^{\infty} e^{-\beta U(x)}
dx}$ are each (approximately) normalized Boltzmann probability
distributions for the two states, where $\beta \equiv \frac{1}{kT}$ \footnote{We neglect the
momentum of the particle in the statistical calculations. In these cases
it decouples from the position and thus has no bearing on the final results.}.
It should be emphasized that the probability distribution of \sc one\rm s and \sc zero\rm s 
is not determined by the thermodynamics of the system; it is externally imposed by 
the computation. 

The Gibbs statistical entropy of such a set of bits is defined to be
$S_G = -\int_{-\infty}^{\infty} \rho(x) \ln \rho(x) dx$.
In the \sc reset-to-one \rm operation, the entropy change of the 
system for the case where the potential is
symmetric about the origin and $p_0^s = p_1^s = 1/2$ is 
$\Delta S_G = -k \ln 2$.

Note that it is not the thermodynamic 
entropy that decreases by $k\ln2$, but rather the Gibbs statistical entropy. 
To use this as a bound on the heat dissipated requires care -- the second law 
is a statement about thermodynamic entropy and thus will not suffice. The literature 
in this regard has led to some controversy; some authors \cite{Piecho} 
have simply considered bits that are initially in a global thermal equilibrium between the
two logical states, giving way
to objections \cite{Shenker,Porod} that any realistic bit must be static and thus cannot
be in such an equilibrium. A recent argument 
by Chris Jarzynski \cite{Jarzynski} shows that 
the heat dissipated into the environment is in fact bounded
from below by $-T \Delta S_{G}$.
In this case, the heat dissipated into the environment is indeed at least $kT\ln 2$. 

An important assumption of Landauer's argument, and one
that is implicit throughout nearly all of the literature on this subject \cite{Leff and Rex} without any
serious discussion, is that the physical realization of the bit is symmetric
between the \sc zero \rm and \sc one \rm states \footnote{In his classic 1961 paper,
Landauer briefly mentions that the basic
physical results are the same if the entropy of the \sc zero \rm state is different from that
of the \sc one \rm state, however he does not also consider the case where the average internal
energies are disparate. Piechocinska also briefly considers asymmetries in the bits but does
not fully explore the consequences.}. However, bits are often represented
in physical systems that do not necessarily have complete symmetry
between the two logical states. Relaxing this assumption will show that the oft-made distinction between logical irreversibility
and reversibility can be turned around: asymmetric bits allow for situations in which the energy cost
of information erasure appears solely during the logically reversible step in a computation. 

To prove this, we will consider a specific physical implementation of a general logically irreversible operation
and explicitly analyze its energy cost. First consider the following toy example.
Suppose that we have a bit of information that is represented in a
physical system, again permanently coupled to a thermal bath at temperature $T$,
where \sc zero \rm consists of a single
physical state with energy $U_0 = 0$ and \sc one \rm consists of four
distinct physical states, each with energy $U_1=2kT\ln2$. Suppose further
that there exists a large potential barrier between the \sc zero \rm
and \sc one \rm states so that the bit is static over any time-scale
of interest. 

The following \sc reset-to-one \rm operation, performed on a set of bits with probability
distribution  $p_0^s = p_1^s = 1/2$, is dissipationless on average.
Step 1: Instantly remove the barrier and allow the system to
equilibrate with the bath. Step 2: Perform work to quasi-statically increase the energy of the
\sc zero \rm state to an arbitrarily large final energy $U_f \gg
kT$. Step 3: Leaving a large barrier like the original one, instantly reduce the energy
of the \sc zero \rm state back to zero. 

The barrier removal is instantaneous and requires no work; since the 
average change in the internal energy of the bits is also zero, the first law implies that 
step one is dissipationless on average. Step two is a thermodynamically reversible 
transformation and hence the heat dissipated is $Q_2 = -T\Delta S_2$, 
where $\Delta S_2$ is the change in entropy in step two. Using the first
law again for step three, $ Q_3 = 0$
because the work performed on the system and the average change
in internal energy are both zero. Therefore the total heat dissipated in this \sc reset-to-one
\rm operation, averaged over the distribution of \sc zero\rm s and
\sc one\rm s, is $\langle Q \rangle_{total} = -T \Delta S_2 = 0$.

A similar calculation shows that $\Delta S_G$ is also
zero. Observe, however, that to return to the initial state of $1/2$ \sc one\rm s and $1/2$ \sc zero\rm s
through a logically reversible operation such as \sc not \rm will be fundamentally dissipative. 

Subsequent to Landauer's distinction between logically irreversible and reversible
operations, a number of researchers showed that all computations could be
performed using solely reversible logic gates \cite{Toffoli,Bennett73}. The recurring theme throughout all
of these implementations of reversible logic is the use of extra bits to store additional
information about the inputs; a closer look at the preceding example reveals 
an analogous situation.

Suppose that the four states in the \sc one \rm configuration of the previous example
are separated from each other by potential barriers and that the precise physical state of the
bit after the \sc reset-to-one \rm operation depends on the input.
In this case the operation is in some sense logically reversible, for a closer examination of the
output will reveal the input. Just as with reversible logic gates, the asymmetry provides the 
extra scratch space to store the necessary additional information. Without the extra space, this additional
information  would leak out into the non-information-bearing degrees of freedom
in the environment and result in heat dissipation.

In reality though, the barriers between the four states need not exist and thus the local thermalization
of the \sc one \rm state implies that this information truly does get destroyed without any 
average change in entropy of the system or environment. The advantage of this outlook is in
demonstrating the necessity of this additional scratch space  
and that its incorporation into the physical structure of the bit can keep the entropy changes isolated 
from the environment. 

Now we will generalize the previous example. 
Suppose that each bit under consideration can be represented as the position of a particle in
an asymmetric one-dimensional bistable potential, $U(x)$, with the \sc zero \rm and \sc one \rm states
as defined earlier. We will set the zero of 
energy to be the average internal energy of the
\sc zero \rm state and let $U_1$ denote the average internal energy of
the \sc one \rm state; $U_b$ will denote the height of the barrier and
$\delta$ its width. As before, the system will be permanently coupled to a thermal bath
at temperature $T$. 

Let $f(\alpha_1,\alpha_2, \ldots, \alpha_n)$ be any arbitrary $n$-input Boolean function,
with the inputs $\{\alpha_i\}$ chosen randomly from a set of bits
with probability distribution $\{p_0^s$, $p_1^s\}$. The physical process carrying out this operation will 
act on $n+1$ bits and will consist of their evolution under a time-dependent Hamiltonian 
$H(\{x_i(t)\},t)$ that first sets the $(n+1)^{th}$ bit to $f$ and then the $n$ input bits to \sc one\rm. 
Here $x_i(t)$ is the position of the particle in the $i^{th}$ bit at time $t$. Let 
$\{p_0^f,p_1^f\}$ be the probability distribution of the bits once the
operation is completed. 

There are generally an infinite number of Hamiltonians that will serve as suitable physical implementations
of the function $f$, but for simplicity we will consider those that add to the $j^{th}$ bit a potential
$V_j(\{x_i\},t)$ with the following properties:

\begin{itemize}
\item [1.] $V_{n+1} = 0$ for $t < 0$ and $t > \tau$.

\item [2.] At a short time $\delta t$ after the operation begins, 
the effect of $V_{n+1}(\delta t)$ is to lower the potential barrier separating the two logical states of the bit to $U_b'$.
This will be an instantaneous change in the limit $\delta t \rightarrow 0$.

\item [3.] As $t \rightarrow (\tau - \delta t$), $V_{n+1}$ slowly raises the potential of either the \sc zero \rm 
or \sc one \rm side depending on the $n$ input bits. In the limit that $\tau \rightarrow \infty$, this is a 
thermodynamically reversible transformation.
 
\item [4.] In the remaining time, $V_{n+1} \rightarrow 0$, which is also an instantaneous change in the limit
$\delta t \rightarrow 0$. 

\item [5.] For all other $j$, the evolution of $V_j$ is similar to that of $V_{n+1}$ except that it begins 
at time $t = \tau$ and always raises the potential of the \sc zero \rm states once the barrier is lowered.
 
\end{itemize}

Note that this procedure is entirely memoryless: the Hamiltonian at time $t$ does not depend on the
earlier states of the bits and hence the information about the inputs truly is erased if $f$ is a logically 
irreversible function.  
We will analyze this process in the limit $\tau \rightarrow \infty$ and $\delta t \rightarrow 0$,
and observe that the evolution of the potentials can be divided into three steps. In the first step, 
the potential barrier separating the \sc zero \rm and \sc one \rm states is instantly lowered to $U_b'$. The second
step consists of a thermodynamically reversible transformation that leaves the bit in the \sc one \rm state
$p_1^f$ of the time by raising the potential of the \sc zero \rm states, and in the \sc zero \rm
state $p_0^f$ of the time by raising the potential of the \sc one \rm states. The last step
instantly drops the raised side of the potential back to its initial state, reverting back to the original 
potential before the operation began. 

Let $p_0^e$ and $p_1^e$ denote the probabilities of
being in \sc zero \rm and \sc one\rm, respectively, after the barrier is initially lowered
and the bit allowed to reach thermal equilibrium with the bath. 
Furthermore, let $Z_0 = \int_{-\infty}^0 e^{-\beta U(x)} dx$ and $Z_1 =
\int_0^{\infty} e^{-\beta U(x)} dx$ be the partition functions for the
\sc zero \rm and \sc one \rm states respectively. Finally, $\langle \cdots \rangle$ will 
denote an average over the distribution of \sc one\rm s and \sc zero\rm s.

Step one is an instantaneous change and requires arbitrarily little work if
$U_b$ is large enough and $\delta$ is small enough. Thus the heat dissipated during
this step, $\langle Q \rangle_1$, is the average change in internal energy:
$\langle Q \rangle_1 = -\langle \Delta U \rangle_1 = U_1(p_1^s - p_1^e)$.
Step two is a thermodynamically reversible transformation, thus
the second law implies that the heat dissipated is $\langle Q\rangle_2 = -T\Delta S_2$,
where $\Delta S_2$ is the change in entropy of step two. Using the first
law again for step three, we find that $ \langle Q\rangle_3 = 0$
because this is an instantaneous process where both the work performed on the system and the average change
in internal energy are negligible. Using the well-known result in equilibrium statistical mechanics that 
$F \equiv -kT\ln Z = U - TS$,
where $Z = \int e^{-\beta U(x)} dx$ is the
usual partition function, we can write the average total dissipated heat
during this operation as
$\langle Q \rangle_{total} = -\langle \Delta U \rangle_1 +
\langle \Delta F \rangle_2 - \langle \Delta U \rangle_2$,
where
\begin{equation}
\langle \Delta F \rangle_2 = (-p_0^fkT\ln Z_0 - p_1^fkT\ln Z_1)  + kT\ln
(Z_0 + Z_1).
\end{equation}
The total heat dissipated per bit is therefore
\begin{equation}
\langle Q \rangle_{total} = U_1(p_1^s - p_1^e) + kT \ln
\frac{Z_0 + Z_1}{Z_0^{p_0^f} Z_1^{p_1^f}} - U_1(p_1^f - p_1^e),
\end{equation}
and so for zero dissipation we require that

\begin{equation}
\label{main}
e^{\beta U_1(p_1^f - p_1^s)} = \frac{Z_0+Z_1}{Z_0^{p_0^f} Z_1^{p_1^f}}.
\end{equation}

Since $\frac{Z_1}{Z_0} = \frac{p_1^e}{p_0^e}$, we can rewrite the
right hand side of eqn. (\ref{main}) as $(\frac{p_1^e}{p_0^e})^{-p_1^f} (1 + \frac{p_1^e}{p_0^e})$. 
As a result we are led to study the properties of the following function:

\begin{equation}
f(\eta) = \frac{1+\eta}{\eta^p} \mbox{for $\eta \geq 0$ and $0 \leq p \leq 1$} .
\end{equation}
A simple analysis shows that $\ln f(\eta)$ continuously takes on all values in the set
$[-p\ln p - (1-p)\ln (1-p),\infty)$.
Letting $\eta = \frac{p_1^e}{p_0^e}$ and $\gamma =
e^{\beta (p_1^f - p_1^s)}$, eqn. (\ref{main}) becomes $U_1 \ln \gamma = \ln f(\eta)$. 

Now observe that given any $\gamma \neq 1$, we can set $U_1$ and $p_1^e/p_0^e$ such that
the above equation is satisfied. For the case where $\gamma = 1$, however, 
it cannot be satisfied except for the trivial situation in which 
$p_1^f = 0$ or $1$, which corresponds to 
both the input and output sets of bits being identically \sc one \rm or \sc zero\rm.  

In the case of a \sc reset-to-one \rm operation, not only can the heat dissipated into the 
thermal bath be made arbitrarily small, but so too can the work necessary to carry out the procedure. 
The average work, $\langle W \rangle$,
performed on each bit is simply the average change in free energy during step two; for \sc reset-to-one\rm, this is
$\langle W \rangle = kT\ln (1 + \frac{Z_0}{Z_1})$.
By setting the width of the \sc one \rm well to be much greater than the width of the \sc zero \rm well,
$\frac{Z_0}{Z_1}$ can be made arbitrarily small;  eqn. (\ref{main}) can simultaneously be satisfied if $U_1$ changes
accordingly.

The preceding analysis indicates that there are other possibilities as well. For instance, an erasure process can absorb heat and
cool the environment. Heat must be dissipated only in returning the bits involved in the computation to their initial
statistics. However, it is still reasonable to follow Landauer in attributing
this ultimate energy cost to information erasure. To see this, consider the following write-erase cycle. 
Suppose that we reset to \sc one \rm a sequence of bits with probability distribution
$\{p_0,p_1\}$. If this probability distribution does not correspond to the equilibrium Boltzmann distribution of the bits, 
$\{\frac{Z_0}{Z_0 + Z_1}, \frac{Z_1}{Z_0 + Z_1}\}$, then first we adjust the energy difference between the two states until the 
thermal equilibrium distribution does correspond to the statistical distribution of the bits. This would not dissipate heat, but would
require some amount of work. Then, using the algorithm prescribed earlier, -- lowering the barrier, raising one side and reinserting
the barrier -- the average heat dissipated can be calculated to be

\begin{equation}
\langle Q_{erase} \rangle = -  p_0 T (S_1 - S_0) - k T(p_0 \ln p_0 + p_1 \ln p_1),
\end{equation}
where $S_0$ and $S_1$ are the entropies of the \sc zero \rm and \sc one \rm states, respectively.  
After the erasure, we readjust the energy
difference between the two states to its initial position, again without any additional heat dissipation. 

To now restore the original sequence of bits, we wish to perform a logically reversible write procedure. In this case
the average heat dissipation is $\langle Q_{write} \rangle = - T p_0 (S_0 - S_1)$, and so the net heat dissipation in the 
write-erase cycle is

\begin{equation}
\langle Q_{total} \rangle= -k T (p_0 \ln p_0 + p_1 \ln p_1),
\end{equation}
which is directly related to the information theoretic entropy change in the erasure step. Thus, while the
erasure step itself can be performed without any net heat dissipation or work input, the energy cost that appears in
a complete write-erase cycle is a direct result of the information loss. 

It is difficult to overstate the impact that Landauer's argument has had on our present
understanding of the fundamental limits of computation and the various paradoxes involving the
second law of thermodynamics. Charles Bennett's famous resolution of the Maxwell demon paradox \cite{Bennett87,Bennett82}
points to information erasure as the single phenomenon
that ultimately saves the second law; the reason Szilard's engine \cite{Szilard} cannot turn heat into work,
it has been argued \cite{Landauer89}, is because information must be erased in completing a cycle of operation, 
necessarily dissipating exactly enough heat to save the second law. Earman and Norton's 
counterexample \cite{Earman} to the thesis of reversible measurement is flawed, 
as Bennett \cite{Bennett03} points out,
because it performs a logically irreversible change in the flow of control of data, hence 
dissipating heat. 

In contrast, the arguments made above illustrate that our current understanding of information
erasure requires modification. Landauer correctly attributed an energy cost to erasure, but this
cost can be shuffled around to any step in a sequence of operations that returns the final distribution
of \sc one\rm s and \sc zero\rm s to the original one, regardless of logical reversibility. 
In discussions of Maxwell's demon or Szilard's engine, it is therefore impossible to point to a single step as being the 
fundamentally dissipative one without knowledge of the specific physical system instantiating the bits.

It is a pleasure to thank Jay Deep Sau and Gavin Crooks for many enjoyable and insightful conversations. I
would also like to extend my profound gratitude towards Chris Jarzynski, Wojcieck Zurek, 
Umesh Vazirani, Joel Moore, John Clarke, Peter Shepard, and Ari Siletz for 
helpful discussions and commenting on various drafts of the paper.

\end{document}